\documentclass[groupedaddress, superscriptaddress, citesort]{revtex4}
\usepackage{amssymb, amsmath}
\usepackage[dvips]{graphicx}

\def\be{\begin{equation}}
\def\ee{\end{equation}}
\def\ba{\begin{array}}
\def\ea{\end{array}}

\begin{document}
\baselineskip=18pt

\title {One-Way Deficit of Two Qubit $X$ States}
\author{Yao-Kun Wang}
\affiliation{Institute of Physics, Chinese Academy of Sciences, Beijing 100190, China}
\affiliation{College of Mathematics,  Tonghua Normal University,
 Tonghua, Jilin 134001, China}
 \author{Naihuan Jing}
  \affiliation{School of Mathematical Sciences, South China University of Technology, Guangzhou, Guangdong 510640, China}
 \affiliation{Department of Mathematics, North Carolina State University, Raleigh, NC27695, USA}

 \author{Shao-Ming Fei}
 \affiliation{School of Mathematical Sciences,  Capital Normal
University,  Beijing 100048,  China}
 \affiliation{Max-Planck-Institute for Mathematics in the Sciences, 04103 Leipzig, Germany}
 \author{Zhi-Xi Wang}
 \affiliation{School of Mathematical Sciences,  Capital Normal
University,  Beijing 100048, China}
\author{Jun-Peng Cao}
\affiliation{Institute of Physics, Chinese Academy of Sciences, Beijing 100190, China}
\affiliation{Collaborative Innovation Center of Quantum Matter, Beijing 100190, China}
\author{Heng Fan}
\affiliation{Institute of Physics, Chinese Academy of Sciences, Beijing 100190, China}
\affiliation{Collaborative Innovation Center of Quantum Matter, Beijing 100190, China}

\begin{abstract}
Quantum deficit originates in questions regarding work extraction from quantum systems
coupled to a heat bath [Phys. Rev. Lett. \textbf{89}, 180402 (2002)].
It links quantum correlations with quantum thermodynamics and provides a new standpoint for understanding quantum non-locality.
In this paper, we propose a new method to evaluate the one-way deficit for a class of two-qubit states.
The dynamic behavior of the one-way deficit under decoherence channel is investigated and it is shown that the one-way deficit of the  $X$ states with five parameters is more robust against the decoherence than the entanglement.
\end{abstract}

\maketitle

\section{Introduction}

The quantum entanglement is a resource in quantum information processing such as teleportation\cite{CH2}, super-dense coding\cite{CH}, quantum cryptography\cite{AK},
remote-state preparation\cite{AK2} and so on.
However, there are quantum correlations other than entanglement
which are also useful and has attracted much attention recently \cite{zurek1,1modi,Giorgi,1Streltsov,modi2}.
One remarkable and widely accepted quantity of quantum correlation is quantum discord.
Quantum discord is a measure of the difference between the  mutual information and maximum classical mutual information,
which is generally difficult to calculate even for two qubit quantum system \cite{Ali,Li,chen,shi,Vinjanampathy}.

Other nonclassical correlations besides entanglement and quantum discord have arisen recently. For example, the quantum deficit \cite{oppenheim,horodecki}, measurement-induced disturbance \cite{luo}, geometric discord \cite{luoandfu,dakic}, and continuous-variable discord \cite{adesso,giorda}, see a review \cite{modi2}. Quantum deficit originates in question how to use
nonlocal operation to extract work from a correlated system coupled to a heat bath \cite{oppenheim}.
It is also closely related with other forms of quantum correlations. Oppenheim \emph{et al.} define the work deficit \cite{oppenheim}
\begin{eqnarray}
\Delta\equiv W_{t}-W_{l},
\end{eqnarray}
where $W_{t}$ is the information of the whole system and $W_{l}$ is the localizable information\cite{horodecki2}.
 As with quantum discord, quantum deficit is also equal to the difference of the mutual information and classical deficit \cite{oppenheim2}.  Recently, Streltsov \emph{et al. }\cite{Streltsov0,chuan} give the definition of the one-way information deficit (one-way deficit) by the relative entropy, which reveals an important role of quantum deficit as a resource for the distribution of entanglement. One-way deficit by von Neumann measurement on one side is given by\cite{streltsov}
\begin{eqnarray}
\Delta^{\rightarrow}(\rho^{ab})=\min\limits_{\{\Pi_{k}\}}S(\sum\limits_{k}\Pi_{k}\rho^{ab}\Pi_{k})-S(\rho^{ab}).\label{definition}
\end{eqnarray}
From the definition we can find that the one-way deficit and  quantum discord are exactly different kinds of quantum correlation.
One may wonder whether the analytical formula or the calculation method for a class of two-qubit states like quantum discord
can be obtained. In this paper,
we will endeavor to calculate the one-way deficit for $X$ quantum states with five parameters.

\section{One-Way Deficit for $X$ States with Five Parameters}

We first introduce  the form of two qubit $X$ states. By using proper local unitary transformations,  we can write $\rho^{ab}$ as
\begin{eqnarray}
\rho^{ab}=\frac{1}{4}(I\otimes I+\textbf{r}\cdot\sigma\otimes I+I\otimes\textbf{s}\cdot\sigma+\sum_{i=1}^3c_i\sigma_i\otimes\sigma_i), \label{state1}
\end{eqnarray}
where \textbf{r} and \textbf{s} are Bloch vectors and $\{\sigma_i\}_{i=1}^3$ are
the standard Pauli matrices. When \textbf{r}=\textbf{s}=\textbf{0},
$\rho$ reduces to the two-qubit Bell-diagonal states.
Then, we assume that the Bloch vectors are in $z$ direction,  that is,  $\textbf{r}=(0, 0, r)$,  $\textbf{s}=(0, 0, s)$. The state in Eq. (\ref{state1}) turns into the following form
\begin{eqnarray}
\rho^{ab}=\frac{1}{4}(I\otimes I+r\sigma_{3}\otimes I+I\otimes s\sigma_{3}+\sum_{i=1}^3c_i\sigma_i\otimes\sigma_i),
\end{eqnarray}
In the computational basis ${|00\rangle, |01\rangle, |10\rangle, |11\rangle}$, its density matrix is
\begin{eqnarray}\label{state3}
\rho = \frac{1}{4} \left(
\begin{array}{cccc}
1+r+s+c_3
& 0 & 0 & c_1 -c_2 \\
0 & 1+r-s-c_3 & c_1+c_2 & 0 \\
0 & c_1 +c_2 & 1-r+s-c_3
& 0 \\
c_1 -c_2 & 0 & 0 & 1-r-s+c_3
\end{array}
\right) \,.
\end{eqnarray}
From Eq. (4) in \cite{chen}, after some algebraic calculations, we can obtain that parameters $x, y, s, u, t$ in \cite{chen}
can be substituted for $r, s, c_{1}, c_{2}, c_{3}$ of the $X$ states in Eq. (\ref{state3}) successively and
\begin{eqnarray}\label{canshu1}
r, s, c_{1}, c_{2}, c_{3}\in[-1,1].
\end{eqnarray}

One can also change them to be $x$ or $y$ direction via
an appropriate local unitary transformation without losing its diagonal property of the correlation terms \cite{kim}.

The eigenvalues of the $X$ states in Eq. (\ref{state3}) are given by
\begin{eqnarray}
u_\pm=\frac{1}{4}[1-c_3\pm\sqrt{(r-s)^2+(c_1+c_2)^2} ],\nonumber\\
v_\pm=\frac{1}{4}[1+c_3\pm\sqrt{(r+s)^2+(c_1-c_2)^2} ].\nonumber
\end{eqnarray}

The entropy is given by
\begin{eqnarray}\label{entropy3}
S(\rho)&=&2-[\frac{1}{4}(1-c_3+\sqrt{(r-s)^2+(c_1+c_2)^2})\log(1-c_3+\sqrt{(r-s)^2+(c_1+c_2)^2})\nonumber\\
       & &+\frac{1}{4}(1-c_3-\sqrt{(r-s)^2+(c_1+c_2)^2})\log(1-c_3-\sqrt{(r-s)^2+(c_1+c_2)^2})\nonumber\\
       & &+\frac{1}{4}(1+c_3+\sqrt{(r+s)^2+(c_1-c_2)^2})\log(1+c_3+\sqrt{(r+s)^2+(c_1+c_2)^2})\nonumber\\
       & &+\frac{1}{4}(1+c_3-\sqrt{(r+s)^2+(c_1-c_2)^2})\log(1+c_3-\sqrt{(r+s)^2+(c_1-c_2)^2})]\nonumber\\.
\end{eqnarray}

Next, we evaluate the one-way deficit of the $X$ states in Eq. (\ref{state3}). Let
\begin{eqnarray}
\{\Pi_{k}=|k\rangle\langle k|, k=0, 1\}\nonumber
\end{eqnarray}
be the local measurement for the particle $b$ along the computational base ${|k\rangle}$;  then any von Neumann measurement for the particle $b$ can be written as
\begin{eqnarray}
\{B_{k}=V\Pi_{k}V^{\dag}: k=0, 1\}\nonumber
\end{eqnarray}
for some unitary $V\in U(2)$. For any unitary $V$, we have
\begin{eqnarray}
V=tI+i\vec{y}\vec{\sigma}\nonumber
\end{eqnarray}
with $t\in R$, $\vec{y}=(y_{1}, y_{2}, y_{3})\in R^{3}$, and $t^{2}+y_{1}^{2}+y_{2}^{2}+y_{3}^{2}=1. $
After the measurement ${B_{k}}$, the state $\rho^{ab}$ will be changed to the ensemble $\{{\rho_{k}, p_{k}}\}$ with
\begin{eqnarray}
\rho_{k}: =\frac{1}{p_{k}}(I\otimes B_{k})\rho(I\otimes B_{k})\nonumber\\
p_{k}=tr(I\otimes B_{k})\rho(I\otimes B_{k}).\nonumber
\end{eqnarray}
To evaluate $\rho_{k}$ and $p_{k}$, we write
\begin{eqnarray}
p_{k}\rho_{k}=(I\otimes B_{k})\rho(I\otimes B_{k})=\frac{1}{4}(I\otimes V)(I\otimes \Pi_{k})[I+r\sigma_{3}\otimes I+sI\otimes V^{\dag}\sigma_{3}V^{\dag}+\sum_{j=1}^{3} c_{j}\sigma_{j}\otimes (V^{\dag} \sigma_{j} V)](I\otimes \Pi_{k})(I\otimes V^{\dag}).\nonumber
\end{eqnarray}

By the relations \cite{luo}
\begin{eqnarray}
V^{\dag}\sigma_{1}V=(t^{2}+y_{1}^{2}-y_{2}^{2}-y_{3}^{2})\sigma_{1}+2(ty_{3}+y_{1}y_{2})\sigma_{2}+2(-ty_{2}+y_{1}y_{3})\sigma_{3},\label{condition2} \nonumber \\
V^{\dag}\sigma_{2}V=2(-ty_{3}+y_{1}y_{2})\sigma_{1}+(t^{2}+y_{2}^{2}-y_{1}^{2}-y_{3}^{2})\sigma_{2}+2(ty_{1}+y_{2}y_{3})\sigma_{3},\label{condition3} \nonumber\\
V^{\dag}\sigma_{3}V=2(ty_{2}+y_{1}y_{3})\sigma_{1}+2(-ty_{1}+y_{2}y_{3})\sigma_{2}+(t^{2}+y_{3}^{2}-y_{1}^{2}-y_{2}^{2})\sigma_{3},\label{condition4}\nonumber
\end{eqnarray}
and
\begin{eqnarray}
\Pi_{0}\sigma_{3}\Pi_{0}=\Pi_{0}, \Pi_{1}\sigma_{3}\Pi_{1}=-\Pi_{1}, \Pi_{j}\sigma_{k}\Pi_{j}=0, for j=0, 1, k=1, 2, \label{condition5}\nonumber
 \end{eqnarray}
After some algebra, we obtain
\begin{eqnarray}
p_{0}\rho_{0}=\frac{1}{4}[I+sz_{3}I+c_{1}z_{1}\sigma_{1}+c_{2}z_{2}\sigma_{2}+(r+c_{3}z_{3})\sigma_{3}]\otimes(V\Pi_{0}V^{\dag}),\nonumber\\
p_{1}\rho_{1}=\frac{1}{4}[I-sz_{3}I-c_{1}z_{1}\sigma_{1}-c_{2}z_{2}\sigma_{2}+(r-c_{3}z_{3})\sigma_{3}]\otimes(V\Pi_{1}V^{\dag}),\nonumber
\end{eqnarray}
where
\begin{eqnarray}
z_{1}=2(-ty_{2}+y_{1}y_{3}), \quad z_{2}=2(ty_{1}+y_{2}y_{3}), \quad z_{3}=t^{2}+y_{3}^{2}-y_{1}^{2}-y_{2}^{2}.\label{condition6}\nonumber
\end{eqnarray}

Next, we will evaluate the eigenvalues of $\sum\limits_{k}\Pi_{k}\rho^{ab}\Pi_{k}$ by
\begin{eqnarray}
\sum\limits_{k}\Pi_{k}\rho^{ab}\Pi_{k}=p_{0}\rho_{0}+p_{1}\rho_{1},
\end{eqnarray}
and
\begin{eqnarray}
& &p_{0}\rho_{0}+p_{1}\rho_{1}\nonumber\\
&=&\frac{1}{4}[(I+r\sigma_{3})+(sz_{3}I+c_{1}z_{1}\sigma_{1}+c_{2}z_{2}\sigma_{2}+c_{3}z_{3}\sigma_{3})]\otimes(V\Pi_{0}V^{\dag})\nonumber\\
& &+\frac{1}{4}[(I+r\sigma_{3})-(sz_{3}I+c_{1}z_{1}\sigma_{1}+c_{2}z_{2}\sigma_{2}+c_{3}z_{3}\sigma_{3})]\otimes(V\Pi_{1}V^{\dag})\nonumber\\
&=&\frac{1}{4}(I+r\sigma_{3})\otimes(V\Pi_{0}V^{\dag}+V\Pi_{1}V^{\dag})\nonumber\\
& &+\frac{1}{4}(sz_{3}I+c_{1}z_{1}\sigma_{1}+c_{2}z_{2}\sigma_{2}+c_{3}z_{3}\sigma_{3})\otimes(V\Pi_{0}V^{\dag}-V\Pi_{1}V^{\dag})\nonumber\\
&=&\frac{1}{4}(I+r\sigma_{3})\otimes I+\frac{1}{4}(sz_{3}I+c_{1}z_{1}\sigma_{1}+c_{2}z_{2}\sigma_{2}+c_{3}z_{3}\sigma_{3})\otimes V\sigma_{3}V^{\dag}.\nonumber
\end{eqnarray}

The eigenvalues of $p_{0}\rho_{0}+p_{1}\rho_{1}$  are the same with the states $(I\otimes V^{\dag})(p_{0}\rho_{0}+p_{1}\rho_{1})(I\otimes V)$, and
\begin{eqnarray}\label{value2}
(I\otimes V^{\dag})(p_{0}\rho_{0}+p_{1}\rho_{1})(I\otimes V)=\frac{1}{4}(I+r\sigma_{3})\otimes I+\frac{1}{4}(sz_{3}I+c_{1}z_{1}\sigma_{1}+c_{2}z_{2}\sigma_{2}+c_{3}z_{3}\sigma_{3})\otimes\sigma_{3}.
\end{eqnarray}

The eigenvalues of the equation (\ref{value2}) are

\begin{eqnarray}
\lambda_{1,2}=\frac{1}{4}\left(1-sz_{3}\pm\sqrt{r^{2}-2rc_{3}z_{3}+c_{1}^{2}z_{1}^{2}+c_{2}^{2}z_{2}^{2}+c_{3}^{2}z_{3}^{2}}\right)\nonumber\\
\lambda_{3,4}=\frac{1}{4}\left(1+sz_{3}\pm\sqrt{r^{2}+2rc_{3}z_{3}+c_{1}^{2}z_{1}^{2}+c_{2}^{2}z_{2}^{2}+c_{3}^{2}z_{3}^{2}}\right)
\end{eqnarray}

It can be directly verified that
\begin{eqnarray}
z_{1}^{2}+z_{2}^{2}+z_{3}^{2}=1.\label{condition7}\nonumber
\end{eqnarray}

Set $\phi=z_{3}$, and
\begin{eqnarray}\label{canshu2}
\phi\in[-1,1].
\end{eqnarray}

Let us put $\theta=c_{1}^{2}z_{1}^{2}+c_{2}^{2}z_{2}^{2}+c_{3}^{2}z_{3}^{2}, c=\min\{|c_{1}|, |c_{2}|, |c_{3}|\}, C=\max\{|c_{1}|, |c_{2}|, |c_{3}|\},$ then $c^{2}=\min\{c_{1}^{2}, c_{2}^{2}, c_{3}^{2}\}$, $C^{2}=\max\{c_{1}^{2}, c_{2}^{2}, c_{3}^{2}\}$, then $c^{2} \leq \theta \leq C^{2}$, and the equality can be readily attained by appropriate choice of $t,y_{j}$ \cite{luo}.
Therefore, we see that the range of values allowed for $\theta$ is $[c^{2}, C^{2}]$.

The entropy of $\sum\limits_{k}\Pi_{k}\rho^{ab}\Pi_{k}$ is
\begin{eqnarray}
S(\sum\limits_{k}\Pi_{k}\rho^{ab}\Pi_{k})&=&f(\phi,\theta )=-\sum\limits_{i=1}^{4}\lambda_{i}\log\lambda_{i}\nonumber\\
&=&2-\frac{1}{4}[(1-s\phi+\sqrt{r^{2}-2rc_{3}\phi+\theta})\log(1-s\phi+\sqrt{r^{2}-2rc_{3}\phi+\theta})\nonumber\\
& &+(1-s\phi-\sqrt{r^{2}-2rc_{3}\phi+\theta})\log(1-s\phi-\sqrt{r^{2}-2rc_{3}\phi+\theta})\nonumber\\
& &+(1+s\phi+\sqrt{r^{2}+2rc_{3}\phi+\theta})\log(1+s\phi+\sqrt{r^{2}+2rc_{3}\phi+\theta})\nonumber\\
& &+(1+s\phi-\sqrt{r^{2}+2rc_{3}\phi+\theta})\log(1+s\phi-\sqrt{r^{2}+2rc_{3}\phi+\theta})].\nonumber\\
\end{eqnarray}

From Eq.(\ref{canshu1}), (\ref{canshu2}), we can obtain $1\mp s\phi\geq0$ and
\begin{eqnarray}
\frac{\partial f}{\partial \theta}&=&\frac{1}{\ln[256]}\left[\frac{\ln(1-s\phi-\sqrt{r^{2}-2rc_{3}\phi+\theta})-\ln(1-s\phi+\sqrt{r^{2}-2rc_{3}\phi+\theta})}{\sqrt{r^{2}-2rc_{3}\phi+\theta}}\right.\nonumber\\
& &\left.+\frac{\ln(1+s\phi-\sqrt{r^{2}+2rc_{3}\phi+\theta})-\ln(1+s\phi+\sqrt{r^{2}+2rc_{3}\phi+\theta})}{\sqrt{r^{2}+2rc_{3}\phi+\theta}}\right]\nonumber\\
&=&\frac{1}{\ln[256]}\left[\frac{\ln\frac{1-s\phi-\sqrt{r^{2}-2rc_{3}\phi+\theta}}{1-s\phi+\sqrt{r^{2}-2rc_{3}\phi+\theta}}}{\sqrt{r^{2}-2rc_{3}\phi+
\theta}}+\frac{\ln\frac{1+s\phi-\sqrt{r^{2}+2rc_{3}\phi+\theta}}{1+s\phi+\sqrt{r^{2}+2rc_{3}\phi+\theta}}}{\sqrt{r^{2}+2rc_{3}\phi+\theta}}\right]<0.
\end{eqnarray}
 It converts the problem about $\min S(\sum\limits_{k}\Pi_{k}\rho^{ab}\Pi_{k})$ to the problem about the function of one variable $\phi$ for minimum. That is
\begin{eqnarray}\label{min3}
\min S(\sum\limits_{k}\Pi_{k}\rho^{ab}\Pi_{k})&=&\min\limits_{\phi} f(\phi,C )\nonumber\\
&=&\min\limits_{\phi} \{2-\frac{1}{4}[(1-s\phi+\sqrt{r^{2}-2rc_{3}\phi+C^{2}})\log(1-s\phi+\sqrt{r^{2}-2rc_{3}\phi+C^{2}})\nonumber\\
& &+(1-s\phi-\sqrt{r^{2}-2rc_{3}\phi+C^{2}})\log(1-s\phi-\sqrt{r^{2}-2rc_{3}\phi+C^{2}})\nonumber\\
& &+(1+s\phi+\sqrt{r^{2}+2rc_{3}\phi+C^{2}})\log(1+s\phi+\sqrt{r^{2}+2rc_{3}\phi+C^{2}})\nonumber\\
& &+(1-s\phi-\sqrt{r^{2}+2rc_{3}\phi+C^{2}})\log(1+s\phi-\sqrt{r^{2}+2rc_{3}\phi+C^{2}})]\}.\nonumber\\
\end{eqnarray}

By Eqs. (\ref{definition}), (\ref{entropy3}), (\ref{min3}), the one-way deficit of the $X$ states in Eq. (\ref{state3}) is given by
\begin{eqnarray}
\Delta^{\rightarrow}(\rho^{ab})&=&\min\limits_{\{\Pi_{k}\}}S(\sum\limits_{k}\Pi_{k}\rho^{ab}\Pi_{k})-S(\rho^{ab})\nonumber\\
&=&\frac{1}{4}\left[(1-c_3+\sqrt{(r-s)^2+(c_1+c_2)^2})\log(1-c_3+\sqrt{(r-s)^2+(c_1+c_2)^2})\right.\nonumber\\
       & &+(1-c_3-\sqrt{(r-s)^2+(c_1+c_2)^2})\log(1-c_3-\sqrt{(r-s)^2+(c_1+c_2)^2})\nonumber\\
       & &+(1+c_3+\sqrt{(r+s)^2+(c_1-c_2)^2})\log(1+c_3+\sqrt{(r+s)^2+(c_1-c_2)^2})\nonumber\\
       & &\left.+(1+c_3-\sqrt{(r+s)^2+(c_1-c_2)^2})\log(1+c_3-\sqrt{(r+s)^2+(c_1-c_2)^2})\right]\nonumber\\
       & &-\max\limits_{\phi} \frac{1}{4}\left[(1-s\phi+\sqrt{r^{2}-2rc_{3}\phi+C^{2}})\log(1-s\phi+\sqrt{r^{2}-2rc_{3}\phi+C^{2}})\right.\nonumber\\
& &+(1-s\phi-\sqrt{r^{2}-2rc_{3}\phi+C^{2}})\log(1-s\phi-\sqrt{r^{2}-2rc_{3}\phi+C^{2}})\nonumber\\
& &+(1+s\phi+\sqrt{r^{2}+2rc_{3}\phi+C^{2}})\log(1+s\phi+\sqrt{r^{2}+2rc_{3}\phi+C^{2}})\nonumber\\
& &\left.+(1+s\phi-\sqrt{r^{2}+2rc_{3}\phi+C^{2}})\log(1+s\phi-\sqrt{r^{2}+2rc_{3}\phi+C^{2}})\right],\nonumber\\
\end{eqnarray}
where $C=\max\{|c_{1}|, |c_{2}|, |c_{3}|\}, \phi\in[-1,1].$

For an example, we set $r=0.2, s=0.3, c_{1}=0.3, c_{2}=-0.4, c_{3}=0.56$, and use the minimun command
\begin{eqnarray}
\text{MinValue}[\{\Delta^{\rightarrow}(\rho^{ab}),-1\leq\phi\leq1\},\phi]
\end{eqnarray}
in ``Wolfram Mathematics8.0'' software, and obtain the value of the one-way deficit $0.130614$.

When $r=s=0$, $\rho$ reduces to the two-qubit Bell-diagonal states. One-way deficit of Bell-diagonal states is
\begin{eqnarray}
 \Delta^{\rightarrow}(\rho^{ab})&=&\min\limits_{\{\Pi_{k}\}}S(\sum\limits_{k}\Pi_{k}\rho^{ab}\Pi_{k})-S(\rho^{ab})\nonumber\\
&=&\frac{1}{4}[(1-c_1-c_2-c_3)\log (1-c_1-c_2-c_3)\nonumber\\
& &+(1-c_1+c_2+c_3)\log (1-c_1+c_2+c_3)\nonumber\\
& &+(1+c_1-c_2+c_3)\log (1+c_1-c_2+c_3)\nonumber\\
& &+(1+c_1+c_2-c_3)\log (1+c_1+c_2-c_3)]\nonumber\\
& &-\frac{1-C}{2}\log (1-C)-\frac{1+C}{2}\log (1+C)
\end{eqnarray}
which is in consistent with the result using the simultaneous diagonalization theorem obtained in \cite{wang}.

It is worth mentioning that we have obtained a formula for solving one-way deficit. It is more simpler than the method using the joint entropy theorem\cite{shao}.

\section{\bf Dynamics of one-way deficit under local nondissipative channels}\label{II}

The concurrence of the $X$ states in Eq. (\ref{state3}) can be calculated in terms of the eigenvalues of $\rho\widetilde{\rho}$,  where $\widetilde{\rho}=\sigma_y\otimes \sigma_y\rho^*\sigma_y\otimes \sigma_y$.
The eigenvalues of $\rho\widetilde{\rho}$ are
\begin{eqnarray}
\lambda_{5}&=&\frac{1}{16}(c_1-c_2-\sqrt{(1+c_3)^2-(r+s)^2})^2\nonumber\\
&=&\frac{1}{16}(c_1-c_2-\sqrt{(1+r+s+c_3)(1-r-s+c_3)})^2,\nonumber
\end{eqnarray}
\begin{eqnarray}
\lambda_{6}&=&\frac{1}{16}(c_1-c_2+\sqrt{(1+c_3)^2-(r+s)^2})^2\nonumber\\
&=&\frac{1}{16}(c_1-c_2+\sqrt{(1+r+s+c_3)(1-r-s+c_3)})^2,\nonumber
\end{eqnarray}
\begin{eqnarray}
\lambda_{7}&=&\frac{1}{16}(c_1+c_2-\sqrt{(1-c_3)^2-(r-s)^2})^2\nonumber\\
&=&\frac{1}{16}(c_1+c_2-\sqrt{(1+r-s-c_3)(1-r+s-c_3)})^2,\nonumber
\end{eqnarray}
\begin{eqnarray}
\lambda_{8}&=&\frac{1}{16}(c_1+c_2+\sqrt{(1-c_3)^2-(r-s)^2})^2\nonumber\\
&=&\frac{1}{16}(c_1+c_2+\sqrt{(1+r-s-c_3)(1-r+s-c_3)})^2.\nonumber
\end{eqnarray}
The concurrence of the $X$ states in Eqs. (\ref{state3}) is given by
\begin{widetext}
\begin{eqnarray}
C(\rho^{ab})=\max\{2\max\{\sqrt{\lambda_{5}}, \sqrt{\lambda_{6}}, \sqrt{\lambda_{7}}, \sqrt{\lambda_{8}}\}
-\sqrt{\lambda_{5}}-\sqrt{\lambda_{6}}-\sqrt{\lambda_{7}}-\sqrt{\lambda_{8}}, 0 \}.
\label{twoqubitconcurrence}
\end{eqnarray}
\end{widetext}

In the following we consider that the $X$ states in Eq. (\ref{state3}) undergoes the phase flip channel \cite{Maziero},  with the Kraus operators
$\Gamma_0^{(A)}=$ diag$(\sqrt{1-p/2}, \sqrt{1-p/2})\otimes I$,  $\Gamma_1^{(A)}=$ diag$(\sqrt{p/2}, -\sqrt{p/2})\otimes I$,
$\Gamma_0^{(B)}= I \otimes$ diag$(\sqrt{1-p/2}, \sqrt{1-p/2}) $,  $\Gamma_1^{(B)}= I \otimes$ diag$(\sqrt{p/2}, -\sqrt{p/2}) $,  where $p=1-\exp(-\gamma t)$,  $\gamma$ is
the phase damping rate \cite{Maziero, yu}. Let $\varepsilon(\cdot)$ represent the operator of decoherence. Then under the phase flip channel  we have
\begin{eqnarray}
\varepsilon(\rho)&=& \frac{1}{4}(I\otimes I+r\sigma_{3}\otimes I+I\otimes s \sigma_3+(1-p)^2c_1\sigma_1\otimes\sigma_1\nonumber\\
    &&+(1-p)^2c_2\sigma_2\otimes\sigma_2+c_3\sigma_3\otimes\sigma_3).
\end{eqnarray}

We will only consider the following further simplified family of the $X$ states in Eq. (\ref{state3}), where
\begin{eqnarray}
|c_{1}|<|c_{2}|<|c_{3}|, \label{condition9}
\end{eqnarray}

As $\varepsilon(\rho)$ satisfies conditions in Eqs. (\ref{state3}), (\ref{condition9}) and the one-way deficit of the $\rho^{ab}$ under the phase flip channel is given by
\begin{eqnarray}
\Delta^{\rightarrow}(\varepsilon(\rho^{ab}))
&=&\frac{1}{4}\left[(1-c_3+\sqrt{(r-s)^2+(1-p)^4(c_1+c_2)^2})\log(1-c_3+\sqrt{(r-s)^2+(1-p)^4(c_1+c_2)^2})\right.\nonumber\\
       & &+(1-c_3-\sqrt{(r-s)^2+(1-p)^4(c_1+c_2)^2})\log(1-c_3-\sqrt{(r-s)^2+(1-p)^4(c_1+c_2)^2})\nonumber\\
       & &+(1+c_3+\sqrt{(r+s)^2+(1-p)^4(c_1-c_2)^2})\log(1+c_3+\sqrt{(r+s)^2+(1-p)^4(c_1-c_2)^2})\nonumber\\
       & &\left.+(1+c_3-\sqrt{(r+s)^2+(1-p)^4(c_1-c_2)^2})\log(1+c_3-\sqrt{(r+s)^2+(1-p)^4(c_1-c_2)^2})\right]\nonumber\\
       & &-\max\limits_{\phi} \frac{1}{4}\left[(1-s\phi+\sqrt{r^{2}-2rc_{3}\phi+c_{3}^{2}})\log(1-s\phi+\sqrt{r^{2}-2rc_{3}\phi+c_{3}^{2}})\right.\nonumber\\
& &+(1-s\phi-\sqrt{r^{2}-2rc_{3}\phi+c_{3}^{2}})\log(1-s\phi-\sqrt{r^{2}-2rc_{3}\phi+c_{3}^{2}})\nonumber\\
& &+(1+s\phi+\sqrt{r^{2}+2rc_{3}\phi+c_{3}^{2}})\log(1+s\phi+\sqrt{r^{2}+2rc_{3}\phi+c_{3}^{2}})\nonumber\\
& &\left.+(1+s\phi-\sqrt{r^{2}+2rc_{3}\phi+c_{3}^{2}})\log(1+s\phi-\sqrt{r^{2}+2rc_{3}\phi+c_{3}^{2}})\right].\nonumber\\
\end{eqnarray}

As an example, for $r=0.2$, $s=0.3, c_{1}=0.3, c_{2}=-0.4, c_{3}=0.56$, the dynamic behavior of correlation of the state under the phase flip channel is depicted in Fig.1. Here one sees that the concurrence become zero after the transition. We find that sudden death of entanglement appears at $p=0.217617$. Therefore for these states the entanglement is weaker against the decoherence than the one-way deficit.

\begin{figure}[h]
\scalebox{2.0}{\includegraphics[width=3.25cm]{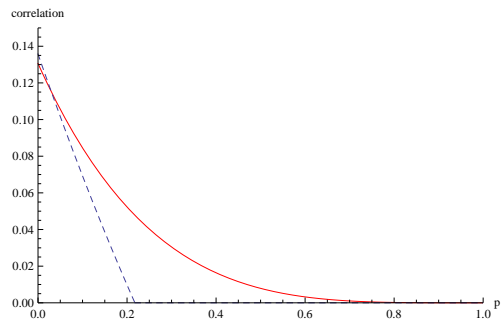}}
\caption{(Color online) Concurrence(blue dashed line) and one-way deficit(red solid line) under phase flip channel for $r=0.2$, $s=0.3$,  $c_1=0.3, $ $c_2=-0.4$ and $c_3=0.56$.}
\label{transition}
\end{figure}

\section{\bf summary}\label{IIII}
We have given a new method to evaluate the one-way deficit for $X$ states with five parameters. By this way, we can evaluate one-way deficit of the wide range states than the method using the simultaneous diagonalization theorem. Meanwhile, this way is more simpler than the method using the joint entropy theorem. The dynamic behavior of the one-way deficit under decoherence channel is investigated. It is shown that one-way deficit of the  $X$ states is more robust against the decoherence than quantum entanglement.

\bigskip
\noindent {\bf Acknowledgments}  This work was supported by NSFC (11175248).


\begin{thebibliography}{18}
\bibitem{CH2} C. H. Bennett, G. Brassard, Cr\'{e}peau, R. Jozsa, A. Peres, and W. K. Wootters, Phys. Rev. Lett. \textbf{70}, 1895 (1993).
\bibitem{CH} C. H. Bennett and S. Wiesner, Phys. Rev. Lett. \textbf{69}, 2881(1992).
\bibitem{AK} A. K. Ekert, Phys. Rev. Lett. \textbf{67}, 661 (1991).
\bibitem{AK2} A. K. Pati, Phys. Rev. A \textbf{63}, 014302 (2000); C. H. Bennett,
D. P. DiVincenzo, P. W. Shor, J. A. Smolin, B. M. Terhal, and W. K. Wootters,
Phys. Rev. Lett. \textbf{87}, 077902 (2001).

\bibitem{zurek1} L. Henderson and V. Vedral, J. Phys. A \textbf{34}, 6899 (2001); H. Ollivier and W. H. Zurek, Phys. Rev. Lett. \textbf{88}, 017901(2002).
\bibitem{1modi} K. Modi, T. Paterek, W. Son, V. Vedral, and M. Williamson, Phys. Rev. Lett. \textbf{104}, 080501 (2010).
\bibitem{Giorgi}G. L. Giorgi, B. Bellomo, F. Galve, and R. Zambrini, Phys. Rev. Lett. \textbf{107}, 190501 (2011).
\bibitem{1Streltsov} A. Streltsov, G. Adesso, M. Piani, and D. Bru{\ss}, Phys. Rev. Lett. \textbf{109}, 050503 (2012).
\bibitem{modi2} K. Modi, A. Brodutch, H. Cable, T. Paterek and V. Vedral, Rev. Mod. Phys. \textbf{84}. 1655 (2012).
\bibitem{Ali} M. Ali, A. R. P. Rau, G. Alber, Phys. Rev. A \textbf{81}, 042105 (2010).
\bibitem{Li} B. Li, Z. X. Wang and S. M. Fei, Phys. Rev. A \textbf{83}, 022321 (2011).
\bibitem{chen} Q. Chen, C. Zhang, S. Yu, X. X. Yi and C. H. Oh, Phys. Rev. A \textbf{84}, 042313 (2011).
\bibitem{shi} M. Shi, C. Sun, F. Jiang, X. Yan and J. Du, Phys. Rev. A \textbf{85}, 064104 (2012).
\bibitem{Vinjanampathy} S. Vinjanampathy, A. R. P. Rau, J. Phys. A \textbf{45}, 095303 (2012).
\bibitem{oppenheim} J. Oppenheim, M. Horodecki, P. Horodecki and R. Horodecki,  Phys. Rev. Lett. \textbf{89},  180402 (2002).
\bibitem{horodecki} M. Horodecki, K. Horodecki, P. Horodecki,  R. Horodecki, J. Oppenheim, A. Sen(De) and U. Sen,  Phys. Rev. Lett.  \textbf{90},  100402 (2003).
\bibitem{luo} S. Luo,  Phys. Rev. A \textbf{77}, 042303 (2008).
\bibitem{luoandfu} S. Luo and S. Fu, Phys. Rev. A \textbf{82}, 034302 (2010).
\bibitem{dakic} B. Dakic, V. Vedral, and C. Brukner, Phys. Rev. Lett. \textbf{105},190502 (2010).
\bibitem{adesso} G. Adesso and A. Datta, Phys. Rev. Lett. \textbf{105}, 030501 (2010).
\bibitem{giorda} P. Giorda and M. G. A. Paris, Phys. Rev. Lett. \textbf{105}, 020503 (2010).
\bibitem{horodecki2} M. Horodecki, P. Horodecki, R. Horodecki, J. Oppenheim, A. Sen(De), U. Sen and B. Synak, Phys. Rev. A \textbf{71}, 062307 (2005); M. Horodecki, P. Horodecki and J. Oppenheim, Phys. Rev. A \textbf{67}, 062104 (2003).
\bibitem{oppenheim2} J. Oppenheim, K. Horodecki, M. Horodecki, P. Horodecki and R. Horodecki, Phys. Rev. A \textbf{68},  022307 (2003).
\bibitem{Streltsov0} A. Streltsov, H. Kampermann and D. Bru{\ss}, Phys. Rev. Lett. \textbf{108}, 250501 (2012).
\bibitem{chuan} T. K. Chuan, J. Maillard, K. Modi, T. Paterek, M. Paternostro and M. Piani, Phys. Rev. Lett. \textbf{109}, 070501 (2012).
\bibitem{streltsov} A. Streltsov, H. Kampermann and D. Bru{\ss},  Phys. Rev. Lett. \textbf{106},  160401 (2011).
\bibitem{kim}  H. Kim,  M. R. Hwang,  E. Jung and D. Park,  Phys. Rev. A \textbf{81},  052325 (2010).
\bibitem{wang} Y. K. Wang, T. Ma, B. Li, Z. X. Wang, Commun. Theor. Phys \textbf{59}, 540 (2013).
\bibitem{shao} L. H. Shao, Z. J. Xi, Y. M. Li, Commun. Theor. Phys \textbf{59}, 285 (2013).
\bibitem{Maziero} J.~Maziero,  L.~C. C{\'e}leri,  R.~M. Serra and V.~Vedral,  Phys. Rev. A \textbf{80},  044102
(2009).
\bibitem{yu} T. Yu and J. H. Eberly,  Phys. Rev. Lett. \textbf{97}, 140403 (2006).

\end{thebibliography}
\end{document}